\documentclass[a4paper, pra, twocolumn, superscriptaddress,showpacs]{revtex4}

\usepackage{color}
\usepackage{soul}

\usepackage{amssymb}
\usepackage{amsmath}

\usepackage{epsfig}
\usepackage{color}
\usepackage{graphics, graphicx}
\usepackage{bbold}
\usepackage{psfrag}
\usepackage{mathcomp}
\usepackage{subfigure}
\def\beq{\begin{equation}}
\def\eeq{\end{equation}}
\def\bea{\begin{eqnarray}}
\def\eea{\end{eqnarray}}

\def\be{\begin{equation}}
\def\ee{\end{equation}}

\begin{document}

\author{A. Privitera}

\affiliation{Institut f\"ur Theoretische Physik, Johann Wolfgang Goethe-Universit\"at, 60438 Frankfurt am Main, Germany}
 \affiliation{Dipartimento di Fisica, Universit\`a di Roma La Sapienza, I-00185 Roma, Italy}
\affiliation{Democritos National Simulation Center, Consiglio Nazionale delle Ricerche, Istituto Officina dei Materiali (CNR-IOM) 
and International School for Advanced Studies (SISSA), I-34136 Trieste, Italy} 
\author{I. Titvinidze}
\affiliation{Institut f\"ur Theoretische Physik, Johann Wolfgang Goethe-Universit\"at, 60438 Frankfurt am Main, Germany}
\author{S.-Y. Chang}
\affiliation{Institute for Quantum Optics and Quantum Information of the Austrian Academy of Sciences, and Institute for Theoretical Physics,
                     University of Innsbruck, A-6020 Innsbruck, Austria}
\affiliation{Department of Physics, The Ohio State University, Columbus, OH 43210, USA}
\author{S. Diehl}
\author{A. J. Daley}
\affiliation{Institute for Quantum Optics and Quantum Information of the Austrian Academy of Sciences, and Institute for Theoretical Physics,
                     University of Innsbruck, A-6020 Innsbruck, Austria}
\author{W. Hofstetter}
\affiliation{Institut f\"ur Theoretische Physik, Johann Wolfgang Goethe-Universit\"at, 60438 Frankfurt am Main, Germany}

\date{\today}

\title{Loss-induced phase separation and pairing for 3-species atomic lattice fermions}

\begin{abstract}
We study the physics of a three-component Fermi gas in an optical lattice, in the presence of a strong three-body constraint arising due to three-body loss.
Using analytical and numerical techniques, we show that an atomic color superfluid phase is formed in this system and undergoes phase
separation between unpaired fermions and superfluid pairs. This phase separation survives well above the critical temperature,
 giving a clear experimental signature of the three-body constraint. 
\end{abstract}
\maketitle

The recent realization of multi-component degenerate Fermi mixtures in experiments \cite{lithium,ae} offers exciting new opportunities
to observe physics beyond that conventionally studied for two-component fermions in the solid state. Three-species mixtures can, 
e.g., give rise to a competition between bound three-body trions and an atomic color superfluid (ACS) phase, 
where unpaired fermions coexist with superfluid (SF) pairs of different components
\cite{acstheory,cherng07}.
The current limitation for observing this physics in three-component Lithium gases is the high rate of three-body loss due to a 
three-body Efimov resonance in this system \cite{lithium, lithiumresonance}. However, loading the gas into an optical lattice could
be used to suppress losses, as a large rate of onsite three-body loss can prevent coherent tunneling processes from populating any
site with three particles \cite{daley_bosons,cirac}, as was demonstrated for the case of two-body losses in Feshbach molecules
\cite{twobody}. For three-body loss, this mechanism would provide an effective three-body hard-core constraint
\cite{daley_bosons,diehl09,kantian}, suppressing the actual loss events. 

\begin{figure}[t]
\includegraphics[width=90mm,angle=-0]{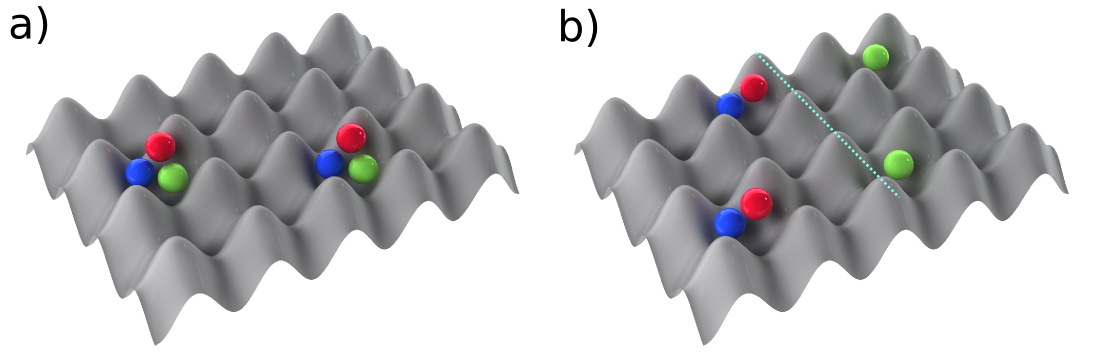}
\caption{(Color online) Sketch of the phases (a) without and (b) with the three-body constraint induced by losses.
The trionic phase expected in the strong-coupling regime is replaced by a phase-separated ACS.}
\label{Fig1} 
\end{figure}

Here we address the key question of how this loss-induced constraint affects the many-body physics in a lattice 
in 2D and 3D. We focus on a regime of medium-to-strong attractive interactions with pronounced asymmetries in
the interactions between different species, corresponding to the regime of strong losses in current experiments with $^6$Li \cite{lithiumresonance}.  On a lattice, we find that the loss-induced constraint not only inhibits trion formation, but also drives the homogeneous ACS phase towards phase separation of SF pairs and unpaired
fermions (see Fig.~\ref{Fig1}). In contrast to two-species Fermi 
mixtures with attractive interactions \cite{capone}, phase separation takes place here even for a number-balanced mixture.
Moreover, this separation persists well above the ACS critical temperature, leaving a strong many-body signature 
of the constraint accessible at current experimental temperatures. 
 
 Below we discuss the model and establish a simple analytical picture for the phase
 separation based on deriving an effective low-energy Hamiltonian at strong-coupling.
 We then analyse the system quantitatively for realistic parameters using Dynamical Mean-field Theory (DMFT)
 and Variational Monte Carlo (VMC) techniques, and present experimental signatures.

\textit{Model} -- A three-component Fermi gas loaded into the lowest Bloch band of an optical lattice can be
 described by the Hamiltonian ($\hbar\equiv 1$)
\begin{equation}\label{eq:CHam}
H\!=\!-J \sum_{\langle i,j\rangle,\sigma}\hat{c}^\dag_{i,\sigma}\hat{c}_{j,\sigma}\! 
-\!\sum_{\sigma,i} \mu_\sigma \hat n_{i,\sigma}\!+\!\!\sum_{\sigma<\sigma^\prime ,i}
\!U_{\sigma,\sigma^\prime}\hat n_{i,\sigma}\hat n_{i,\sigma^\prime},
\end{equation}
where $\hat{c}_{i,\sigma}^\dag$ is a fermionic creation operator for an atom on site $i$ in internal state 
$\sigma =1, 2, 3$, $J$ is the tunnelling amplitude between nearest neighbour sites $\langle i,j \rangle$,  $U_{\sigma,\sigma^\prime}$ are the pairwise onsite interaction
energies, and $\hat n_{\sigma,i}= \hat{c}^\dag_{\sigma,i}\hat{c}_{\sigma,i}$.
 The chemical potentials $\mu_\sigma$ fix the filling factors $n_\sigma$. This model assumes that $J,
 U_{\sigma,\sigma^\prime} \ll \nu$,  where $\nu$ is the energy separation to higher Bloch bands.
 In the presence of strong on-site three-body loss with a rate $\gamma_3$,
 tunnelling processes populating sites with three atoms are suppressed as $J^2/\gamma_3$ \cite{daley_bosons}.
 We thus obtain a model with an effective three-body
 constraint, $\hat n_{i,1}\hat n_{i,2}\hat n_{i,3}\equiv 0$, as discussed in Refs.~\cite{daley_bosons,diehl09,kantian}.

\textit{Parameters for Lithium} -- The requirement $\gamma_3\gg J$ is fulfilled if we take $^6$Li atoms at a magnetic
 field of $500$ Gauss as realized in Refs.~\cite{lithium}. From experimentally measured scattering lengths 
and loss rate constants \cite{lithium}, we obtain for an isotropic cubic lattice in 
3D of depth $8E_R$ in each dimension (where $E_R$ is the recoil energy), that 
$U_{1,2} \approx -41.6J,\, U_{1,3} \approx -25.1J\, U_{2,3} \approx -6.2J$, and 
$\gamma_3\sim 100 J$ \cite{kantian}. Variations in the lattice depth correspond to rescaling 
the values of the interactions given above by a common factor $\lambda$, i.e. $U_{\sigma,\sigma^\prime} \to
\lambda U_{\sigma,\sigma^\prime}$. Throughout this work we focus on a globally {\it{balanced}} mixture, $n_\sigma=n$, in the medium-to-strong coupling regime $(\lambda \in [0.3,1])$.
For comparison of 3D and 2D results, we keep $U_{\sigma,\sigma^\prime}/W$ constant, where $W=4DJ$ is
the width of the lowest Bloch band in $D$ dimensions. 

\textit{Simple picture of pairing and phase separation} -- To establish an analytical understanding, we first choose $\lambda=1$ and 
$n_\sigma = n=1/6$. In this regime the system exhibits clear energy scale separation, (i) being interaction dominated $|U_{\sigma,\sigma^\prime}|/J\gg 1$ 
and (ii) having strongly anisotropic interactions.

We can see how this scale separation leads to pairing and phase separation by generalizing a recently developed technique 
for treating onsite constrained models to fermionic systems \cite{diehl09}. The constrained fermionic theory is exactly mapped onto
a theory in single fermion operators  $t_{\sigma,i}$ and a vector boson $b_{\sigma,i}$ made up of fermions of species 
$\sigma+1,\sigma+2$ on site $i$ (with cyclic property $\sigma + 3 = \sigma$) \cite{diehl_long} without additional constraints. 
Due to the balanced mixture and strong scale separation, with $U_{1,2}$ the largest energy scale, all atoms from the most strongly
interacting combination $\sigma=1,2$ will form pairs, while those with $\sigma=3$ remain unpaired.
Trion formation is suppressed by the constraint. At low energies we obtain the {\it{effective}} model 
$H \approx H_{{\rm eff}}=H_{3F} + H_{BF} + H_{3B}$ \cite{diehl_long}, with

\begin{eqnarray}\label{eq:EffHam}
H_{3F} &=& J \sum_{i,\lambda} \nabla_{i,\lambda} t^{\dag}_{3,i} \nabla_{i,\lambda} t_{3,i}, \ \ 
H_{BF} =J \sum_{\langle i,j \rangle} \tilde n_{3,i} \tilde m_{3,j}, \nonumber \\
H_{3B} &=& -(2J^2/|U_{1,2}|) \sum_{\langle i,j\rangle}\big[ 2  b^{\dag}_{3,i}  b_{3,j}- \tilde m_{3,i}  \tilde m_{3,j}  \big] ,
\end{eqnarray}
where the lattice gradient is defined as $\nabla_{i,\lambda} t_{3,i} \equiv t_{3,i+\mathbf{e}_\lambda} - t_{3,i}$, 
$\mathbf{e}_\lambda$ is the unit vector in the primitive lattice directions, $\tilde m_{\sigma,i}=b_{\sigma,i}^\dag b_{\sigma,i}$
and $H_{3B}$ for the bound state is obtained by integrating out the fermions (1,2)
in perturbation theory. The effective Hamiltonian $H_{{\rm eff}}$ provides us with an asymptotic strong-coupling picture 
in terms of a Bose-Fermi mixture, where fast fermions of the unpaired species with hopping 
$J$ strongly interact with slow bosonic pairs, whose effective hopping parameter is given by $4J^2/|U_{1,2}|\approx J/10 \ll J$. 
The strong (off-site) boson-fermion \textit{repulsion} $J$ in $H_{BF}$ is a \textit{direct consequence} of the three-body 
constraint. This is the term that gives rise to phase separation - however, this scale is of the same order as the fermionic 
hopping and phase-separation is not generic for Bose-Fermi mixtures \cite{bosefermi1}, so the problem must be treated carefully.

We assess the 
instability of the homogeneus system against phase separation by studying the compressibility matrix 
$M_{\sigma\sigma^\prime} = \partial^2 E/\partial n_\sigma \partial n_{\sigma^\prime}$, where $n_\sigma,n_{\sigma^\prime} =n_3,m_3$ and $E$ is the canonical energy for Eq. \eqref{eq:EffHam}. For low filling and at mean-field level, we obtain
$E^{(D)}(n_{3},m_{3}) \approx \pi^2 \frac{W}{D+2} \Big(\frac{D}{\Omega_D}\Big)^{2/D} \, n_3^{2/D + 1}+
W^2/(4D|U_{1,2}|)  m_{3}^2 + W m_{3}n_{3}$, where $\Omega_D = 2\pi,4\pi$ for $D=2,3$. 
Consequently the system will be stable provided $1 \leq A=\pi^2 (D/\Omega_D)^{2/D}D^{-3} \frac{W}{|U_{1,2}|} n_3^{2/D -1 }$. 
For $n_3=1/6$, $A=0.07 (0.11)$ in 3D(2D), and thus the homogeneous system is unstable against phase separation.
Note that there is a critical density below which the homogeneous mixture
is stable in 3D ($n \approx 10^{-4}$ for this setup). In 1D the system is not expected to phase separate on typical 
experimental scales, where the constraint helps instead to stabilize the homogeneous ACS phase \cite{kantian}. 

%
Since in strong-coupling $T_c=\mathcal O(J^2/|U_{1,2}|) \ll J$ \cite{micnas},
 we expect phase separation (with a typical energy scale of the boson-fermion repulsion $J$)
 to be present well above the critical temperature for the disappearance of the SF phase. Finally, as the fermion-boson interaction
is repulsive, the instability will result in spatially separated domains of fermions and
bosons in order to minimize the bulk energy, as shown numerically below.

{\it Numerical approaches} -- In order to treat the system quantitatively in a wider range of parameters and go beyond the effective
description above, we employ two complementary numerical approaches.  Details on both methods
are provided in \cite{symm_long}.

In 3D, we employ DMFT as a 
non-perturbative approach where the (quasi-) momentum dependence of the 
self-energy is neglected, i.e. $\Sigma(\bf{k},\omega)=\Sigma(\omega)$\cite{revdmft}.
We generalized DMFT to the case with three species in order to describe ACS and trionic phase,
 and also investigated the system at
finite temperature. As a solver for DMFT, we used Exact Diagonalization \cite{krauth}. Onsite thermodynamic observables, e.g. densities $n_\sigma=\langle \hat{c}^\dagger_\sigma 
\hat{c}_\sigma \rangle$, double occupancies $d_{\sigma,\sigma^\prime}=\langle \hat{n}_\sigma \hat{n}_{\sigma^\prime}\rangle$ 
and the SF order parameter $P=\langle \hat{c}_1 \hat{c}_2 \rangle$, are estimated directly as averages on the impurity site of the 
auxiliary Anderson model at self-consistency \cite{revdmft}.

The momentum distribution $n_\sigma({\bf{k}})$ and spectral properties
are obtained from the normal components of the single-particle Green function
$G_\sigma({\bf{k}},i\omega_n)$, which is a natural output in DMFT.
 Then, $n_\sigma({\bf{k}})=T\sum_n G_\sigma({\bf{k}},i\omega_n) e^{-i\omega_n 0^-}$
and the average kinetic energy per lattice site is given by 
$K=\frac{1}{{\mathcal{N}}}\sum_{{\bf{k}},\sigma}\varepsilon_{\bf{k}} n_\sigma({\bf{k}})$,
 where $\mathcal{N}$ is the number of lattice sites. 
The internal energy per site $E$ is then $E=K+V_{pot}$, where $V_{pot}=\sum_{\sigma < \sigma^\prime} U_{\sigma, \sigma^\prime} d_{\sigma,\sigma^\prime}$
is the mean potential energy per site.

In 2D, we work with a VMC method, based on a strong-coupling expansion of Eq. \eqref{eq:CHam} \cite{symm_long}.
We can determine the stability of the ACS phase with respect to trion formation and its instability towards phase
 separation by comparing trial wavefunctions for the ground state. For the normal phase we use 
\beq
|NFF \rangle = {\cal J}{\cal P}_3 {\cal P}_D \prod_\sigma \prod_{\varepsilon_{\bf{k},\sigma} \le \varepsilon_{F,\sigma}} \hat{c}_{{\bf k_\sigma},\sigma}^\dagger|0\rangle,
\label{eqn_su3}
\eeq
where ${\cal P}_D$ is a fixed number projection, ${\cal P}_3$ is a projector enforcing the three body constraint, and ${\cal J}$ is a Jastrow variational factor that can also include off-site trion correlations.
For the homogeneous ACS phase we use
\beq
|ACS \rangle  =  {\cal J}{\cal P}_3 {\cal P}_D \prod_{{\bf k}}\left[u_{\bf k} + v_{\bf k} 
\hat{c}_{-{\bf k},1}^\dagger \hat{c}_{{\bf k},2}^\dagger \right] 
\prod_{\varepsilon_{{\bf{k}}',3} \le \varepsilon_{F,3}}\hat{c}_{{\bf k}',3}^\dagger|0 \rangle,
\label{eqn_su2u1}
\eeq
where $u_{\bf k}$ and $v_{\bf k}$ are the usual BCS functions. 

Since DMFT and VMC work respectively in the grandcanonical and canonical ensemble, they provide us with 
complementary descriptions of the system.  This is particularly relevant in the presence of phase separation.

\textit{Numerical results at $T=0$} -- 
We first treat the strong-coupling regime ($\lambda=1$), and characterize the ground state properties
 as a function of the overall density $n$ for fixed interactions. 
In particular, we establish the instability of the homogeneous ACS phase beyond the range of validity 
of the effective Hamiltonian approach and characterize {\it{quantitatively}} the equilibrium mixture.

Within DMFT we find no values of the chemical potentials $\mu_\sigma$ such that the system is in a single homogeneous
phase with the required densities $n_\sigma=n$. This gives evidence for phase separation, in agreement with the general
arguments above. This scenario, in which a mixture of phases has lower ground state energy $E$ than a single homogeneous
phase at $T=0$, is signalled in the grandcanonical framework by the existence of multiple solutions with the same grand 
potential per lattice site $\Omega=E-\sum_\sigma \mu_\sigma n_\sigma$ for given $\mu_\sigma$.

We considered several possible scenarios to build the equilibrium mixture, combining phases with pairing in
 different channels \cite{cherng} and also explicitly suppressing superfluidity.  We found that the equilibrium 
mixture at $T=0$ consists of a fraction $\alpha$ of a two-species SF involving the species with the strongest interaction, 
$1$--$2$, at densities $n^{SF}_1=n^{SF}_2 \equiv n_p, n^{SF}_3=0$, and a fraction $1-\alpha$ of a normal phase which 
accommodates the remaining unpaired species, i.e. $n^{N}_1=n^{N}_2=0, n^{N}_3\equiv n_u$.  The energy of the mixture 
per lattice site is given by $E_{mix}=\alpha E^{SF}(n_p) +(1-\alpha)E^{N}(n_u)$ and $\alpha=n/n_p=1-n/n_u$. Note that a similar
 scenario (though with a different $\alpha$) also applies for much smaller $\lambda$, where the simple analytical picture above 
does not apply. 
Moreover, we find that the densities in the two phases in equilibrium 
are very different, being $n_p \approx 0.75$ and $n_u \approx 0.2$ for $\lambda=1$ and $n=0.16$,
since a (strongly-interacting) two-species SF is in equilibrium with a spin-polarized (and therefore non-interacting) Fermi gas.

\begin{figure}[tb]
\includegraphics[width=80mm,angle=-0]{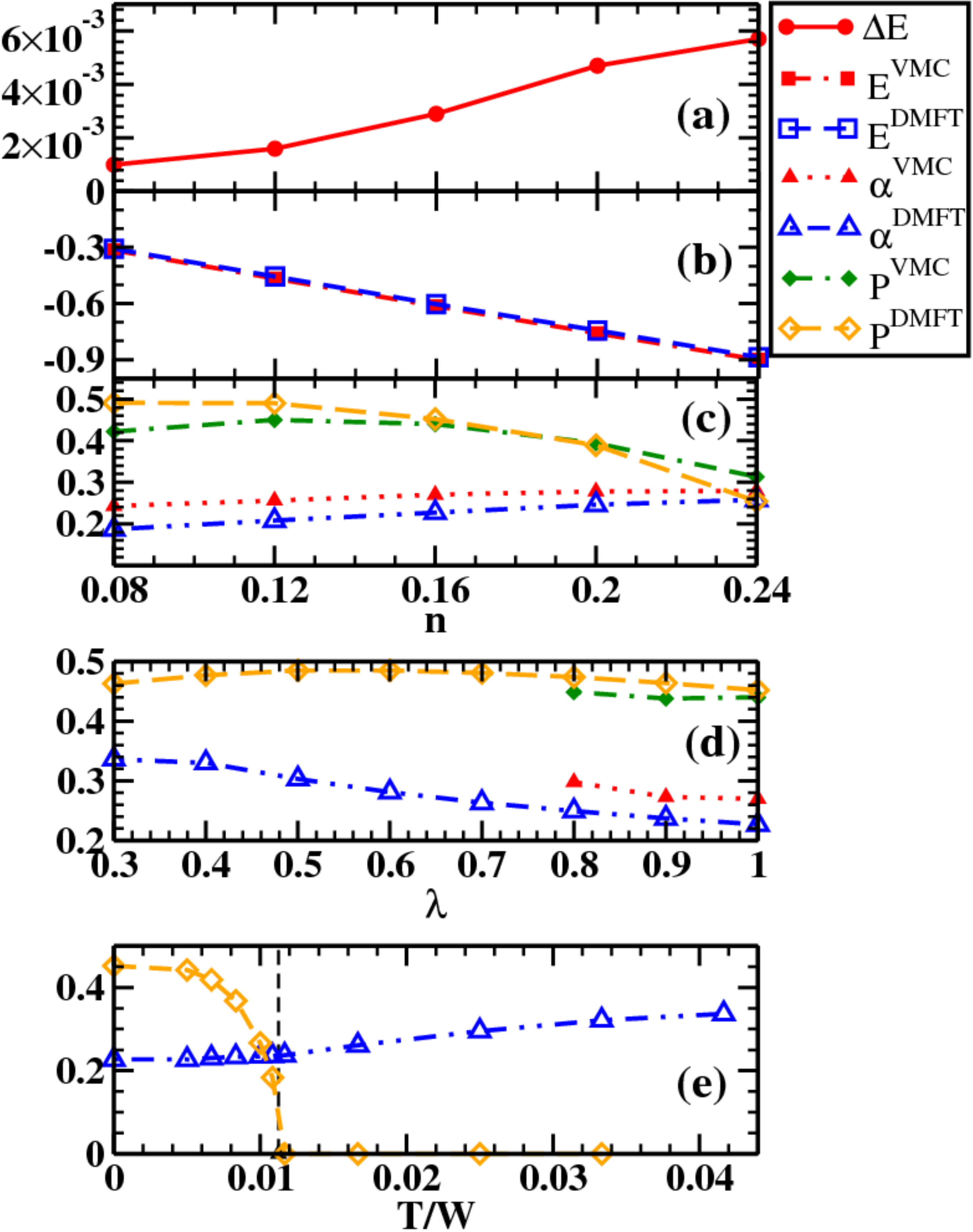}
\caption{(Color online) (a--c) VMC and DMFT results at $T=0$, for $\lambda=1$ 
as a function of $n$: a) Energy gain $\Delta E$ from phase-separation within VMC.
b) Ground state energy $E_{mix}$ of the equilibrium mixture.
c) Fraction of superfluid phase $\alpha$ and superfluid order parameter $P$ at equilibrium.
d) Same as (c), but for $n=0.16$ as a function of $\lambda$.
e) Same as (c), but for $\lambda=1, n=0.16$ at finite $T$. 
Energies are given in units with $W=1$. }
\label{Fig2} 
\end{figure}

This scenario is confirmed via VMC in 2D, where we use an \emph{explicit} Maxwell construction. We find that the energy $E_{mix}$
of the SF plus normal mixture described above is lower than the energy $E_{hom}$ of the homogeneous ACS solution for the same overall 
fixed densities $n_\sigma=n$. We plot $\Delta E=E_{hom}-E_{mix} > 0$ in Fig.~\ref{Fig2}a, where it is evident that the energy difference
between the homogeneus solution and the equilibrium mixture is a very small fraction of $E_{mix}$ shown in Fig.~\ref{Fig2}b. The relative
energy gain from phase separation is therefore very small, probably because the kinetic energy of the unpaired fermions and the 
boson-fermion repulsion are of the same order $\sim J$ (see Eq. \ref{eq:EffHam}). Once rescaled with the bandwidth $W=4DJ$, 
the DMFT and VMC values for the energy $E_{mix}^{DMFT}$ and $E_{mix}^{VMC}$ are in good agreement.
The approximately linear scaling of $E_{mix}$ with the density ($E_{mix} \approx n U_{1,2}$) occurs because the interaction energy 
in the superfluid is much larger than the kinetic energy in the superfluid or normal fluid, and $d_{1,2}\approx n_p$ in the superfluid
phase at strong-coupling. 

To further characterize the equilibrium mixture, we also evaluated the SF order parameter
$P = \lim_{{\bf{R}} \to \infty} P({\bf{R}}) =\sqrt{\frac{1}{\mathcal{N}_{SF}} \sum_i \langle c_{i,2}^\dagger
 c_{i,1}^\dagger c_{i+{\bf{R}},1}c_{i+{\bf{R}},2}}\rangle$, where $\mathcal{N}_{SF}$ is the number of lattice sites
in the SF domain.
As shown in Fig.~\ref{Fig2}c, $\alpha$ increases monotonically with the density in both $2D$
and $3D$. Conversely, $P$ is maximum whenever $n_p=0.5$, similarly to a two-component SF. 

In Fig.~\ref{Fig2}d we show that the phase-separated scenario persists in the intermediate-coupling 
regime where the effective Hamiltonian description fails, rescaling the interactions with a factor $\lambda$
down to $\lambda = 0.3$.

\textit{Numerical results at finite temperature} --
To connect with experiments, we investigated the effect of finite temperature $T$ within DMFT.  For finite 
$T$ the grandcanonical potential is given by  $\Omega=E-TS-\sum_\sigma \mu_\sigma n_\sigma$, where the entropy $S$ is computed 
via the Maxwell relation ${\partial n_\sigma}/{\partial T}={\partial S}/{\partial \mu_\sigma}$. For increasing $T$ the order 
parameter $P$ for the superfluid component decreases and eventually vanishes at $T=T_c$, as shown in Fig. ~\ref{Fig2}e. However,
phase separation persists for $T>T_c$, where both components are normal, as clearly seen from the mixing 
fraction $\alpha$. As explained above, this effect is due to the loss-induced constraint and scale separation. 
This is in contrast to (i) strong coupling in the absence of a constraint, where the trionic phase is favored and 
(ii) the case of symmetric interactions $U_{\sigma,\sigma^\prime}=U$, where phase separation is favoured due to a
 condensation energy gain in the ACS phase and disappears above the $T_c$ \cite{symm_long}. 

\begin{figure}[tb]
\begin{center}
    \begin{tabular}{c}
     \resizebox{70mm}{!}{\includegraphics[angle=-90]{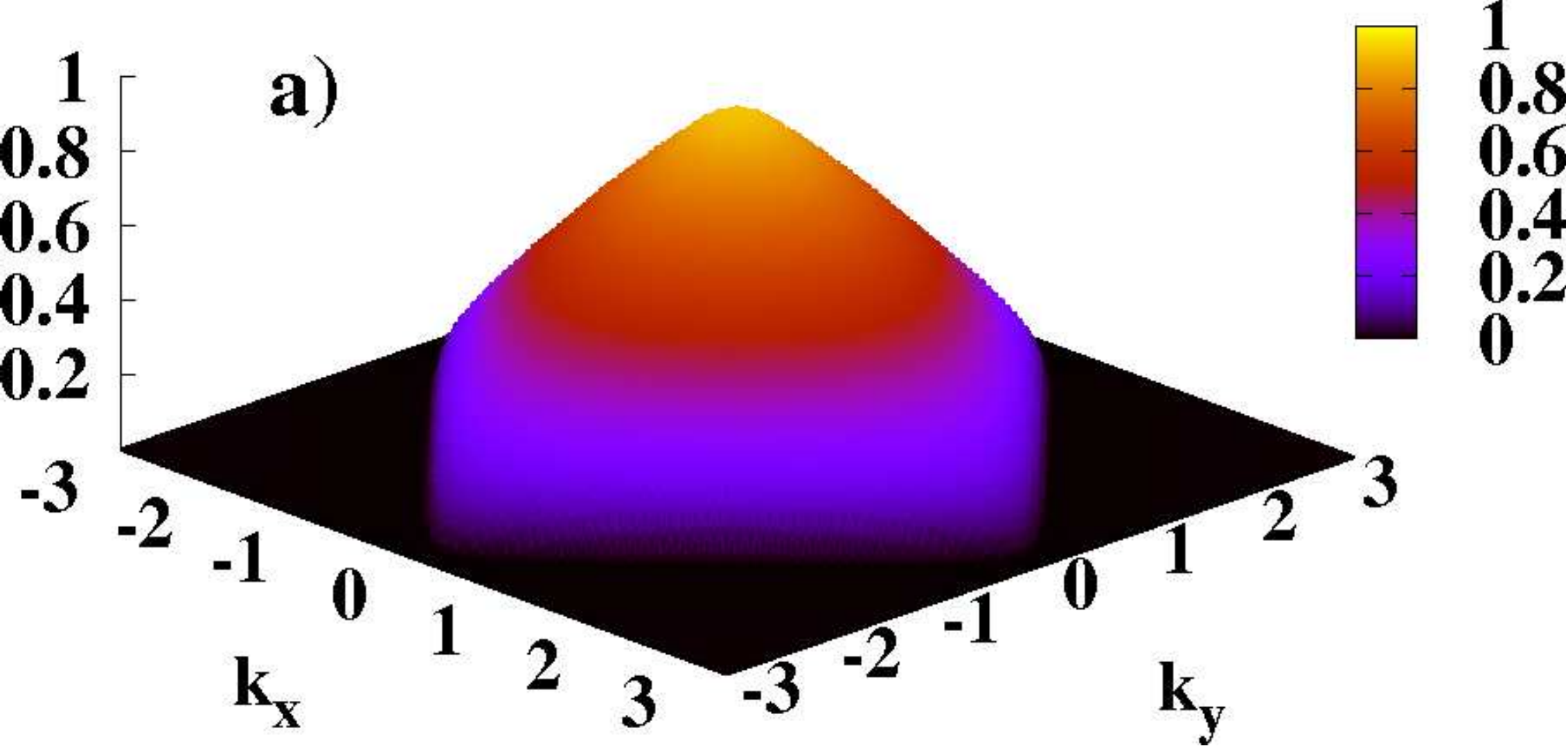}} \\
     \resizebox{70mm}{!}{\includegraphics[angle=-90]{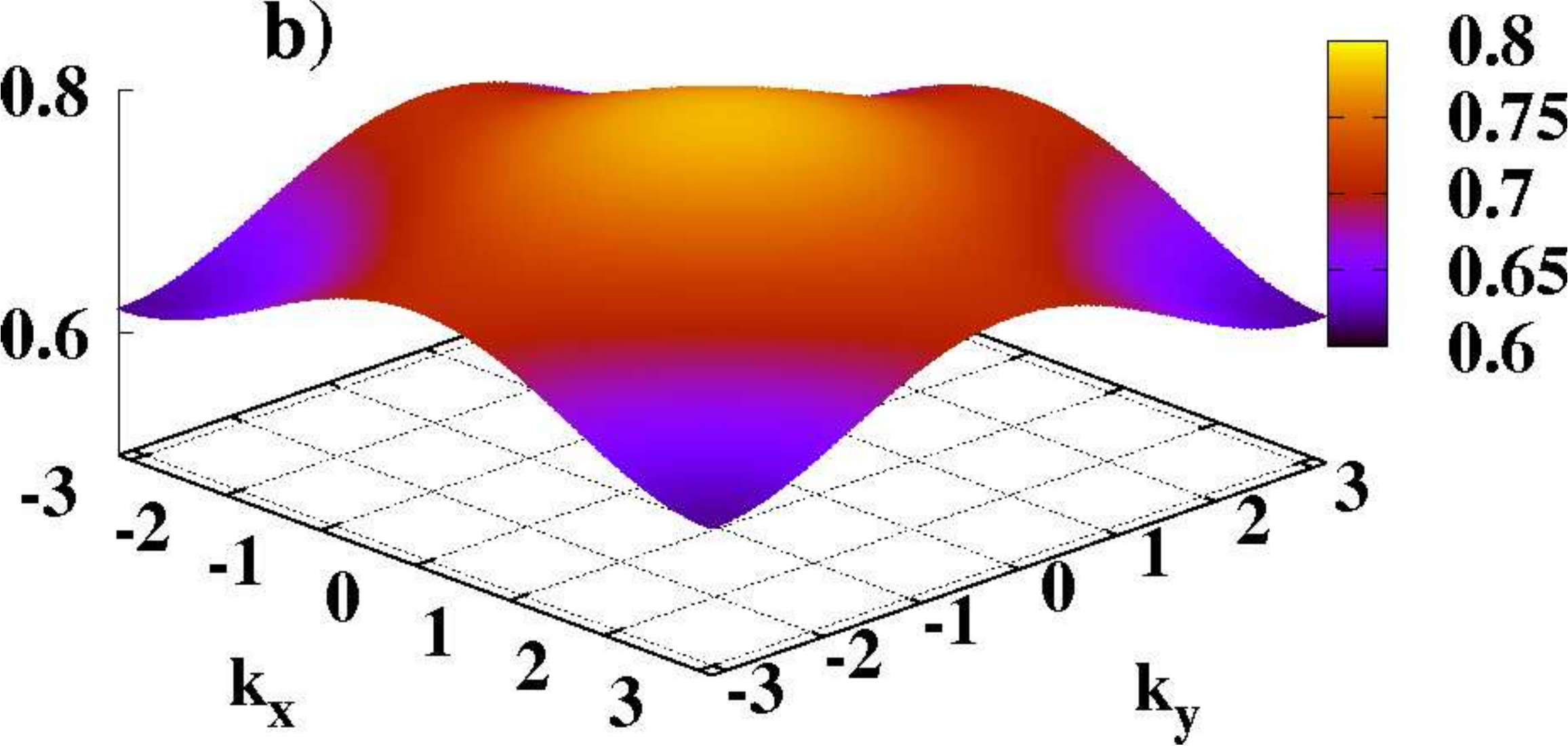}} \\
\hspace{-0.7cm}\resizebox{70mm}{!}{\includegraphics{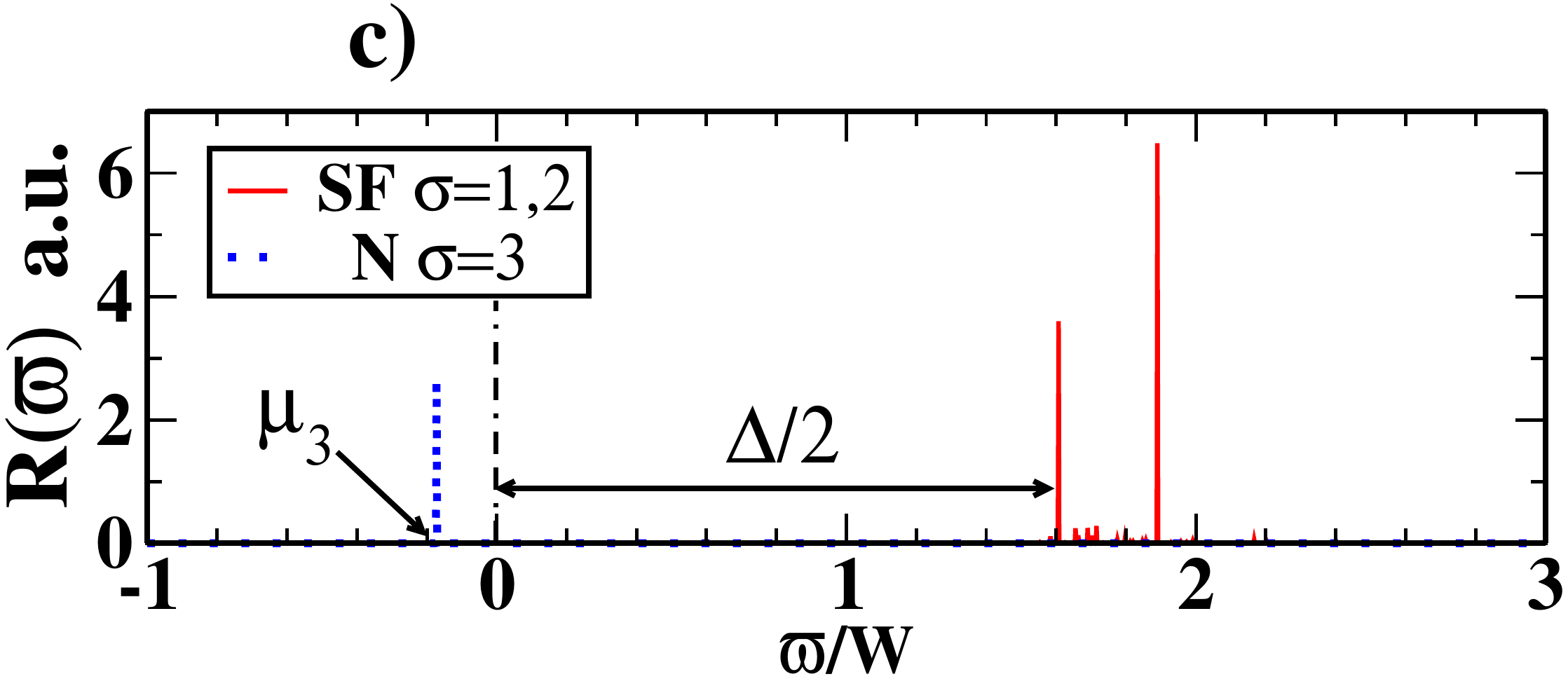}} \\ 
    \end{tabular}
\end{center}
\caption{(Color online) (a,b) $z$-Integrated momentum distribution $\bar{n}_\sigma(k_x,k_y)$ from DMFT for (a) $\sigma=3$ in the normal phase
and (b) $\sigma=1$ (identical to $\sigma=2$) in the SF phase. (c) Raman response $R(\bar{\omega})$ for the normal (dotted blue line) and SF component
(solid red line) of the mixture. All results are for $\lambda=1$, $n=0.16$ and $T=0$.}
\label{Fig4} 
\end{figure} 

\textit{Experimental signatures} -- Clear experimental signatures of this 
scenario appear in the quasi-momentum distribution of the mixture, accessible in state-selective
 time-of-flight (TOF) measurements. Within DMFT we estimated the $z$-integrated momentum 
distribution $\bar{n}_\sigma(k_x,k_y)=\int \tfrac{dk_z}{2\pi} n_\sigma({\bf{k}})$ of the 
two phases in 3D at $T=0$. For the normal phase of the unparied species (Fig.~\ref{Fig4}a),
 this coincides with the Fermi distribution at $T=0$ ($Z=1$,
where $Z$ is the quasiparticle weight). Despite smoothing due to the integration
along the $z$ axis, $\bar{n}$ vanishes identically at large ${\bf{k}}$. If the system were in a balanced ACS phase,
the jump at the Fermi surface of the unpaired species would be strongly renormalized at strong coupling ($Z<1$) and some
particles would be promoted to momenta outside the Fermi surface. The SF component (Fig.~\ref{Fig4}b) shows the weak dependence
of $n_{{\bf{k}}}$ on $k$ as expected for a two-color SF in the BEC regime.

Complementary signatures can be obtained using Raman spectroscopy \cite{bernier}, which allows us
to distinguish different scenarios via the spectral gap. Within DMFT, the Raman
response at zero transferred momentum $R_\sigma(\bar{\omega})$, where $\bar{\omega}=\omega+\mu-\varepsilon_{out}^0$ 
and $\varepsilon^0_{out}$ is the energy of the final state, is given at $T=0$ 
by $R_\sigma(\bar{\omega})=C\sum_{\bf{k}}^{\varepsilon_{\bf{k}} < \bar{\omega}} 
A_\sigma({\bf{k}},\varepsilon_{\bf{k}}-\bar{\omega})$, where $C$ is a constant and
$A_\sigma({\bf{k}},\omega)=-1/\pi {\rm Im} G_\sigma({\bf{k}},\omega+i0^+)$.
As shown in Fig.~\ref{Fig4}c, $R(\bar{\omega})$ for the SF component of the mixture shows a gap 
$\Delta \approx 3.2 W =\mathcal{O} (U_{12})$, while for the normal (non-interacting) component 
$R(\bar{\omega})\propto \delta(\bar{\omega}-\mu_3)$. For a balanced ACS, the Raman peak at 
$\bar{\omega}=\mu_3$ would be broadened due to the interactions and in a trionic phase every component would 
be gapped with $\Delta_\sigma=\mathcal{O}(U_{\sigma,\sigma+1}+U_{\sigma,\sigma+2})$.


We thank M. Baranov and S. Jochim for valuable discussions. Work in Frankfurt is supported
by the German Science Foundation DFG via Sonderforschungsbereich SFB-TRR 49. Work in Innsbruck
 is supported by the Austrian Science Fund through SFB F40 FOQUS and EUROQUAM\_DQS (I118-N16).


\begin{thebibliography}{99}
\bibitem{cherng} In the asymmetric case, only pairing between the natural colours is allowed, as shown in 
Ref.~\cite{cherng07}.
\end{thebibliography}
\end{document}